\documentclass[review]{elsarticle}

\usepackage{lineno,hyperref}
\usepackage{multicol}
\usepackage{CJKutf8}
\modulolinenumbers[5]

\journal{Chinese Physics C}
\usepackage{graphicx}
\usepackage{amsmath}

\begin{document}
\begin{CJK*}{UTF8}{gbsn}
\begin{frontmatter}

\title{Neutron stars including the effects of chaotic magnetic fields and the anomalous magnetic moments}



\author[nju]{Fei Wu(吴飞)}
\ead{howar6hill@yahoo.com}
\author[siap]{Chen Wu(吴琛)}
\ead{wuchenoffd@gmail.com}
\author[nju,nlha,kavli]{Zhong-Zhou Ren(任中洲)}
\ead{zren@nju.edu.cn}
\address[nju]{Department of Physics, Nanjing University, Nanjing 210093, China}
\address[siap]{Shanghai Institute of Applied Physics, Chinese Academy of Science, Shanghai 201800, China}
\address[nlha]{Center of Theoretical Nuclear Physics, National Laboratory of Heavy-Ion Accelerator, Lanzhou 730000, China}
\address[kavli]{Kavli Institute for Theoretical Physics China, CAS, Beijing 100190, China}

\begin{abstract}
In this work the relativistic mean field (RMF) FSUGold model extended to include hyperons is employed to study the properties of neutron stars with strong magnetic fields. The chaotic magnetic field approximation is utilized. The effect of the anomalous magnetic moments (AMMs) is also investigated. It is shown that the equation of state (EOS)of neutron star matter is stiffened by the presence of the magnetic field, which increases the maximum mass of neutron star by around 6\%. The AMMs only have a small influence on the EOS of neutron star matter, and increase the maximum mass of neutron star by 0.02 M$_{sun}$. Neutral particles are spin polarized due to the presence of the AMMs.\\

\end{abstract}



\end{frontmatter}


\section{Introduction}
The magnetic field is one of the most important constituents of the cosmic space and one of the main sources of the dynamics of interacting matter in the universe. Compact stars under strong magnetic fields have drawn much attention. For instance, some authors have evaluated quasinormal modes of massive scalar field of the Ernst spacetime describing a black hole immersed in a uniform magnetic field \cite{brito2014superradiant,konoplya2008quasinormal,wu2015decay}. Further more, there exists very strong magnetic fields in neutron stars. Observational evidence suggests that the  magnetic field strength on the surface of soft gamma repeaters and
anomalous X-ray pulsars could be as high as $10^{14} - 10^{15}$ G \cite{duncan1992formation, thompson1995soft, usov1992millisecond, 1538-4357-486-2-L129, 1538-4357-519-2-L139, kouveliotou1998x}. It exceeds the critical field strength $B_c^e=4.414 \times 10^{13}$ G \cite{broderick2000equation}, thus is expected to influence the properties of neutron stars significantly. Fields larger than $1 \times 10^{18}$ G are expected in the interior of neutron stars due to the scalar viral theorem. The macroscopic properties, such as mass and radius, will depend sensitively on the EOS of strongly magnetized neutron stars. Therefore, it is necessary to study the effect of strong magnetic field on the properties of neutron stars.
 	
 	The energy-momentum tensor of magnetic field is anisotropic. Strictly speaking, the TOV equation \cite{oppenheimer1939massive, wu2011strange} is not valid in this situation. To solve this problem, Lopes and Menezes recently proposed a chaotic magnetic field approximation \cite{lopes2015magnetized}, which is able to avoid this issue. It has been used successfully to study the properties of quark stars \cite{lopes2015magnetized}. Following this line of thought, in our present work, the chaotic magnetic field approximation is extended to investigate the effect of the AMMs on the properties of neutron stars. In Ref. \cite{broderick2000equation}, the contributions of proton and neutron AMMs to hadronic EOS were calculated for the first time. It demonstrated that it is possible for the AMMs to have a significant influence on the EOS of neutron star matter. Later it was generalized to include the contribution from the eight light baryons \cite{broderick2002effects}. Additionally, in most previous works \cite{bandyopadhyay1997quantizing, rabhi2009quark, menezes2009quark, ryu2010medium, rabhi2011warm, mallick2011possible, lopes2012influence, casali2014hadronic, menezes2014repulsive, mallick2014deformation} , a number density-dependent magnetic field is used. A energy density-dependent magnetic field \cite{lopes2015magnetized} proposed recently has got little attention so far. One of our tasks in this work is to study the influence of the energy density-dependent magnetic field.

 	RMF theory has been widely used to study the interaction between baryons and meson fields, since it greatly decreases the complexity of the problem, and has achieved much success\cite{Yuan}. Within the framework of RMF, Todd-Rutel and Piekarewicz  recently proposed the FSUGold model \cite{todd2005neutron}, which is able to reproduce the properties of nuclear matter successfully \cite{piekarewicz2007validating}. Recently this model is utilized to study the EOS of neutron star matter\cite{fattoyev2010sensitivity}. However, the effect of magnetic field is neglected in previous works that include hyperons. In this paper we will use the FSUGold model extended to include hyperons to investigate the properties of neutron stars. Both of the effects of the magnetic field and the AMMs will be considered. The chaotic magnetic field approximation and energy density-dependent magnetic field model will be utilized.

	This work is organized as follows. First we introduce the theoretical framework. Next we study the effect of the magnetic field on the EOS,  mass-radius relation and particle density of neutron stars. Then the effect of the AMMs on these properties is discussed. Finally, some conclusions are drawn.

\section{Theoretical framework}
To describe the EOS of hadronic matter, we employ the RMF theory, in which baryons interact via
the exchange of $\sigma$, $\omega$ and $\rho$ mesons. The baryons under consideration include nucleons and hyperons first investigated by Glendenning \cite{glendenning1982hyperon}. The Lagrangian density of the FSUGold model reads \cite{wu2011strange}
 \begin{align}
\mathcal{L} = &\sum_B
\bar{\psi}_B[i\gamma^{\mu}\partial_\mu-q_B\gamma^{\mu}A_{\mu}-m_{B}+g_{\sigma
B}\sigma \nonumber \\ &
-g_{\omega B}\gamma^\mu\omega_\mu - g_{\rho B}
\gamma^\mu \vec{\tau}\cdot \vec{\rho}^\mu-\frac{1}{2}\mu_N\kappa_B\sigma_{\mu\nu}F^{\mu\nu}
]\psi_B  \nonumber \\ &
+\frac{1}{2}\partial_{\mu}\sigma\partial^{\mu}\sigma -\frac{1}{2}m_\sigma^2 \sigma^2 -
\frac{\kappa}{3!}(g_{\sigma N} \sigma)^3 -\frac{\lambda}{4!}(g_{\sigma N} \sigma)^4 \nonumber \\ &
-\frac{1}{4}\Omega_{\mu\nu}\Omega^{\mu\nu}+\frac{1}{2}m_{\omega}^2\omega_\mu\omega^\mu+
\frac{\zeta}{4!}(g_{\omega N}^2 \omega_\mu \omega^\mu)^2  \nonumber\\ &
-\frac{1}{4}\vec{G_{\mu\nu}}\vec{G^{\mu\nu}}+\frac{1}{2}m_{\rho}^2\vec{\rho_\mu} \cdot \vec{\rho^\mu}
 + \Lambda_v (g_{\rho N}^2 \vec\rho_\mu \cdot \vec\rho^\mu )(g_{\omega N}^2 \omega_\mu \omega^\mu)  \nonumber \\ &
-\frac{1}{4}F_{\mu\nu}F^{\mu\nu}+ \sum_l \bar{\psi}_l[i\gamma^{\mu}\partial_\mu-m_{l}-q_l\gamma^{\mu}A_{\mu}
-\frac{1}{2}\mu_B\kappa_l\sigma_{\mu\nu}F^{\mu\nu}]\psi_l .
\end{align}
The sum in $B$ stands for the entire octet set. $l$ represents  e$^-$ and $\mu^-$. The static properties of these fermions are listed in Table 1. The $g$'s are coupling constants that simulate the strong interaction. $\kappa, \lambda$,  $\zeta$ and $\Lambda_v$ describe the interaction between mesons in the FSUGold model. We list the parameters of the FSUGold model in Table 2. The $m$'s are masses of various particles. $\kappa_B$ denotes the coupling strength for the baryon AMM, and $\mu_N$ is the nuclear magneton. Similarly, $\kappa_l$ denotes the coupling strength for the lepton AMM, and $\mu_B$ is the Bohr magneton. The coupling of the AMM and electromagnetic field is introduced via $\sigma_{\mu\nu}=\frac{i}{2}[\sigma_\mu,\sigma_\nu]$. The mesonic and electromagnetic field tensors take their usual forms:
$\Omega_{\mu\nu}=\partial_\mu \omega_\nu-\partial_\nu \omega_\mu$,
$\overrightarrow{G}_{\mu\nu}=\partial_\mu \overrightarrow{\rho}_\nu-\partial_\nu\overrightarrow{\rho}_\mu$,
$F_{\mu\nu}=\partial_\mu A_\nu-\partial_\nu A_\mu$.


\begin{table}[h]
\centering
\caption{Mass, charge, and coupling strength for the AMM of baryons and leptons considered in this paper \cite{broderick2002effects,mao2003study}.}
\begin{tabular}{cccc}
\hline
Species & Mass(MeV)   & Charge(e)  & coupling strength \\
\hline
p & 938.3 & 1 & 1.79 \\

n & 939.6 & 0 & -1.91 \\

$\Lambda^0$ & 1115.7 & 0 & -0.61 \\

$\Sigma^+$ & 1189.4 & 1 & 1.67 \\

$\Sigma^0$ & 1192.6 & 0 & 1.61 \\

$\Sigma^-$ & 1197.4 & -1 & -0.38 \\

$\Xi^0$ & 1314.8 & 0 & -1.25 \\

$\Xi^0$ & 1321.3 & -1 & 0.06 \\

e$^-$ & 0.51 & -1 & $1.16 \times 10^{-3}$ \\

$\mu^-$ & 105.6 & -1 & $1.17 \times 10^{-3}$ \\
\hline
\end{tabular}

\label{tab:fermions}
\end{table}

\begin{table}[h]
\centering
\caption{Model parameters of the FSUGold model \cite{wu2011strange}.}
\begin{tabular}{cccccccccc}
\hline
$m_\sigma$(MeV) & $m_\omega$(MeV) & $m_\rho$(MeV) & $g_{\sigma N}$ & $g_{\omega N}$ & $g_{\rho N}$ & $\kappa$ & $\lambda$ & $\zeta$ & $\Lambda_v$  \\
\hline
491.5 & 783 & 763 & 10.59 & 14.30 & 11.77 & 1.42 & 0.0238 & 0.06 & 0.03 \\
\hline
\end{tabular}

\label{tab:fsu_para}
\end{table}


To fix the hyperon-meson coupling constants, we take those in the SU(6) quark model for $\rho$ and $\omega$ coupling constants \cite{wu2011strange} :
\begin{eqnarray}
g_{\rho \Lambda} =0, g_{\rho \Sigma} = 2g_{\rho \Xi} =2g_{\rho N}, \nonumber \\
g_{\omega \Lambda} = g_{\omega \Sigma}= 2g_{\omega \Xi} =
\frac{2}{3}g_{\omega N}.
\end{eqnarray}
The $\sigma$ couplings are fixed by fitting hyperon potentials. The obtained couplings are $g_{\sigma \Lambda}=6.31$, $g_{\sigma \Xi}=3.27$, and $g_{\sigma \Sigma}=6.36$ \cite{wu2011strange}.

Within the framework of the RMF theory, meson fields are treated as uniform classical fields. Their equations of motion can be obtained by the application of the action principle \cite{wu2011strange,wu2013neutron,Yuan}. The magnetic field is viewed as an external generated field which has no associated field equation. We also impose $\beta$-equilibrium and charge neutrality conditions to the neutron star matter \cite{broderick2002effects,xu2014superfluidity}.

The main effect of the magnetic field is Landau quantization. The energy spectra for neutron baryons,
charged baryons, and leptons are given by  \cite{rabhi2009quark,broderick2002effects}

 \begin{align}
E^B_{s} =& \sqrt{{k_z}^2+ (\sqrt{{m_B^*}^2+k_x^2+k_y^2}-s\mu_N\kappa_B B)^2} \nonumber \\ &
+ g_{\omega B}\omega +g_{\rho B} \tau_{3B}\rho, \nonumber \\
E^B_{\nu,s} =& \sqrt{{k_z}^2+ (\sqrt{{m_B^*}^2+2\nu|q_B|B}-s\mu_N\kappa_B B)^2} \nonumber \\ &
+ g_{\omega B}\omega +g_{\rho B} \tau_{3B}\rho, \nonumber \\
E^l_{\nu,s} =& \sqrt{{k_z}^2+ (\sqrt{{m_l}^2+2\nu|q_l|B}-s\mu_B\kappa_l B)^2},
 \end{align}
 
where $\nu=0,1,2,3$ ... enumerates the Landau levels of charged fermions; s is +1 for spin-up and
-1 for spin-down cases.
$m_B^*$ is the effective mass of the baryon under consideration. When $\kappa_B$ and $\kappa_l$ are set to zero, the effect of the AMMs is switched off.

The pressure of neutron star matter is obtained from thermodynamic relations at zero temperature \cite{broderick2000equation,rabhi2009quark,Chakrabarty:1996te}
\begin{eqnarray}
P_M=\sum_i \mu_i\rho_i -\epsilon_M,
\end{eqnarray}
where $i$ run over all fermions considered. $\mu_i$ and $\rho_i$ are chemical potential and number density, respectively.
$\epsilon_M$ denotes the energy density of neutron star matter.

To obtain the total energy density and pressure, one must add the contribution of the magnetic field. In the current literature, this is usually done as \cite{rabhi2009quark, menezes2009quark, ryu2010medium, rabhi2011warm, mallick2011possible, lopes2012influence, casali2014hadronic, menezes2014repulsive, mallick2014deformation}
\begin{eqnarray}
\epsilon=\epsilon_M+\frac{B^2}{2}, \nonumber \\
P=P_M+\frac{B^2}{2}.
\end{eqnarray}
However, this seems problematic, since for a magnetic field the energy-momentum tensor is diag($B^2/2, B^2/2, B^2/2, -B^2/2$) \cite{strickland2012bulk}, which is anisotropic. In order to obtain the mass-radius relation of neutron star, the TOV  equation is usually utilized \cite{oppenheimer1939massive, baldo1997microscopic, kalogera1996maximum}. For the TOV equation
to work, the energy-momentum tensor must take the form: diag($\epsilon,P,P,P$). It demands that the energy-momentum tensor be isotropic.

To solve this issue, a chaotic field approximation \cite{lopes2015magnetized} is proposed recently. In this approach, the pressure of magnetic field is $B^2/6$ instead of $B^2/2$. This is consistent with field theory \cite{dexheimer2012hybrid}, in which $P=\frac{1}{3}<T_i^i>$.

Since this approach seems more reasonable, we will adopt it in this paper. Hence, the total energy density and pressure is:
\begin{eqnarray}
\epsilon=\epsilon_M+\frac{B^2}{2}, \nonumber \\
P=P_M+\frac{B^2}{6}.
\end{eqnarray}

The magnetic field varies in the interior of neutron star. But it is still unknown how it varies. In the current literatures, it is usually assumed to be exponential density-dependent \cite{bandyopadhyay1997quantizing, rabhi2009quark, menezes2009quark, ryu2010medium, rabhi2011warm, mallick2011possible, lopes2012influence, casali2014hadronic, menezes2014repulsive, mallick2014deformation}:
\begin{equation}
B=B^{surf}+B_0[ 1-exp(-\beta(\frac{n}{n_0})^\theta ) ],
\end{equation}
where $n$ denotes total number density, $n_0$ is the nuclear saturation density, $B^{surf}$ is the magnetic field on the surface of neutron star, and $B_0$ is a fixed value of the order of the magnetic field in the center of neutron star. $\beta$ and $\theta$ are free parameters.

Since how the magnetic field varies in the interior of neutron star is still unknown, we are free to make other assumption about the variation of
magnetic field. It is the energy density rather than the number density that enters the TOV equation, so it is more natural to let the magnetic field couple to the energy density. We will take this approach, use a model proposed recently in \cite{lopes2015magnetized}:
\begin{eqnarray}
B=B_0(\frac{\epsilon_M}{\epsilon_0})^\gamma+B^{surf},
\end{eqnarray}
where $\epsilon_M$ is the energy density of neutron star matter. $B_0$ is a fixed value of magnetic field, approximately the field strength in the core of neutron star, set to $3.1 \times 10^{18}$ G in this work. $\epsilon_0$ is a fixed value taken to be 5.01 fm$^{-4}$. $B^{surf}$ is the magnetic field on the surface of neutron star, taken as $1 \times 10^{14}$ G. $\gamma$ is a free parameter that can be any positive number.

The magnetic field will cause spin polarization of charged particles, due to the coupling of magnetic moment and magnetic field. The spin polarization has an influence on the superfluidity of neutron stars. It is defined as \cite{dong2013dense}
\begin{eqnarray}
S=\frac{\rho_{\uparrow} -\rho_{\downarrow}}{\rho},
\end{eqnarray}
where $\rho$ denotes the number density of the particle under consideration, $\rho_{\uparrow}$ is the number density of spin-up particle,
while $\rho_{\downarrow}$ is the number density of spin-down particle.

\section{Numerical results}
As pointed out by previous works, the energy density and pressure of neutron star matter is insensitive to magnetic field lower than $1 \times 10^{18}$ G \cite{lopes2015magnetized,gomes2014effects, sinha2013hypernuclear}. Following the approach in Ref. \cite{lopes2015magnetized}, we use a fixed value of magnetic field, namely $3.1 \times 10^{18}$ G, to perform the computation, only taking account of the variation of magnetic field in the calculation of the total energy density and pressure in Eq. (6). This simplification will not change the results considerably.
\subsection{The effect of magnetic field}

First, we study the effect of magnetic field, while ignoring the effect of the AMMs. It is known that the magnetic field can alter the EOS of neutron star matter. In Fig. 1, we plot the EOS without magnetic field and with magnetic field for $\gamma=1, 2, 3, 4, 5$. It shows that the EOS of neutron star matter is stiffened by the magnetic field. The smaller $\gamma$ is, the stiffer the EOS becomes for $\epsilon<6.2$  fm$^{-4}$. This can be understood as follows. The ratio
of pressure to energy density due to magnetic field (equal to 1/3) is larger than that due to matter (less than 1/7) in the  chaotic magnetic field approach. So the stronger the magnetic field is, the stiffer the EOS becomes. For energy density not too large, the magnetic field is stronger for smaller $\gamma$.

\begin{figure}[!h]
\centering
\includegraphics[scale=0.4]{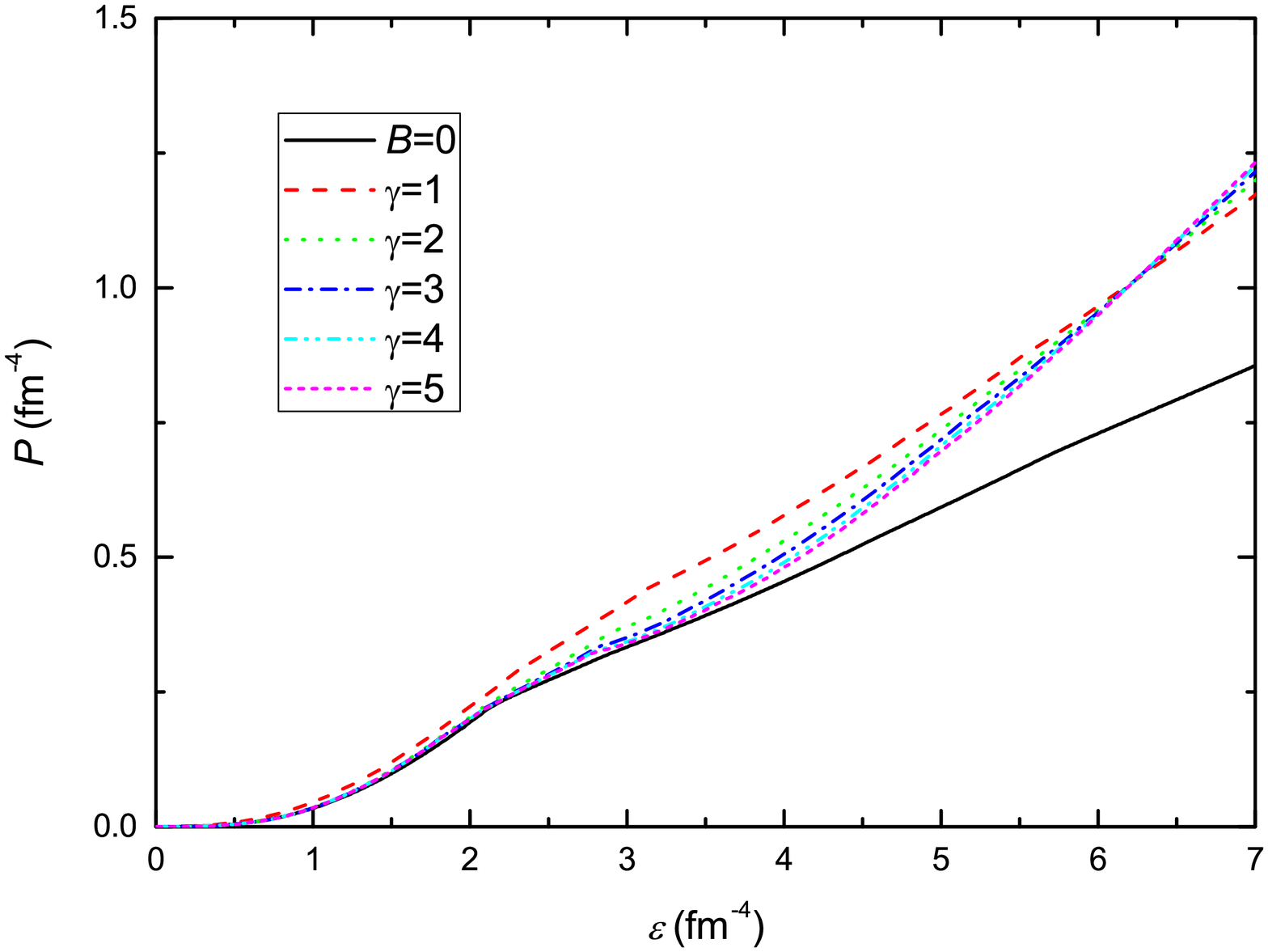}
\caption{ (color online) The EOS of neutron star matter without the magnetic field and with the magnetic field for different $\gamma$.}
\label{fig:s}
\end{figure}

The stiffening of the EOS will increase the maximum mass of neutron star. In Fig. 2, the mass-radius relation of neutron star is displayed.
Indeed, the presence of magnetic field produces an increase in the maximum mass. The smaller $\gamma$ is, the larger the maximum mass and the corresponding radius are.
In Table 3, we list the macroscopic properties of maximum mass neutron stars for various configurations.  For $2\leq \gamma \leq 5$, the maximum mass is all around 1.44 M$_{sun}$, an increase of 6\% from the case without magnetic field. For $2\leq \gamma \leq 5$, the maximum mass neutron star has a radius around 10.8 km, smaller than the field-free case (11.4 km).

\begin{figure}[!h]
\centering
\includegraphics[scale=0.4]{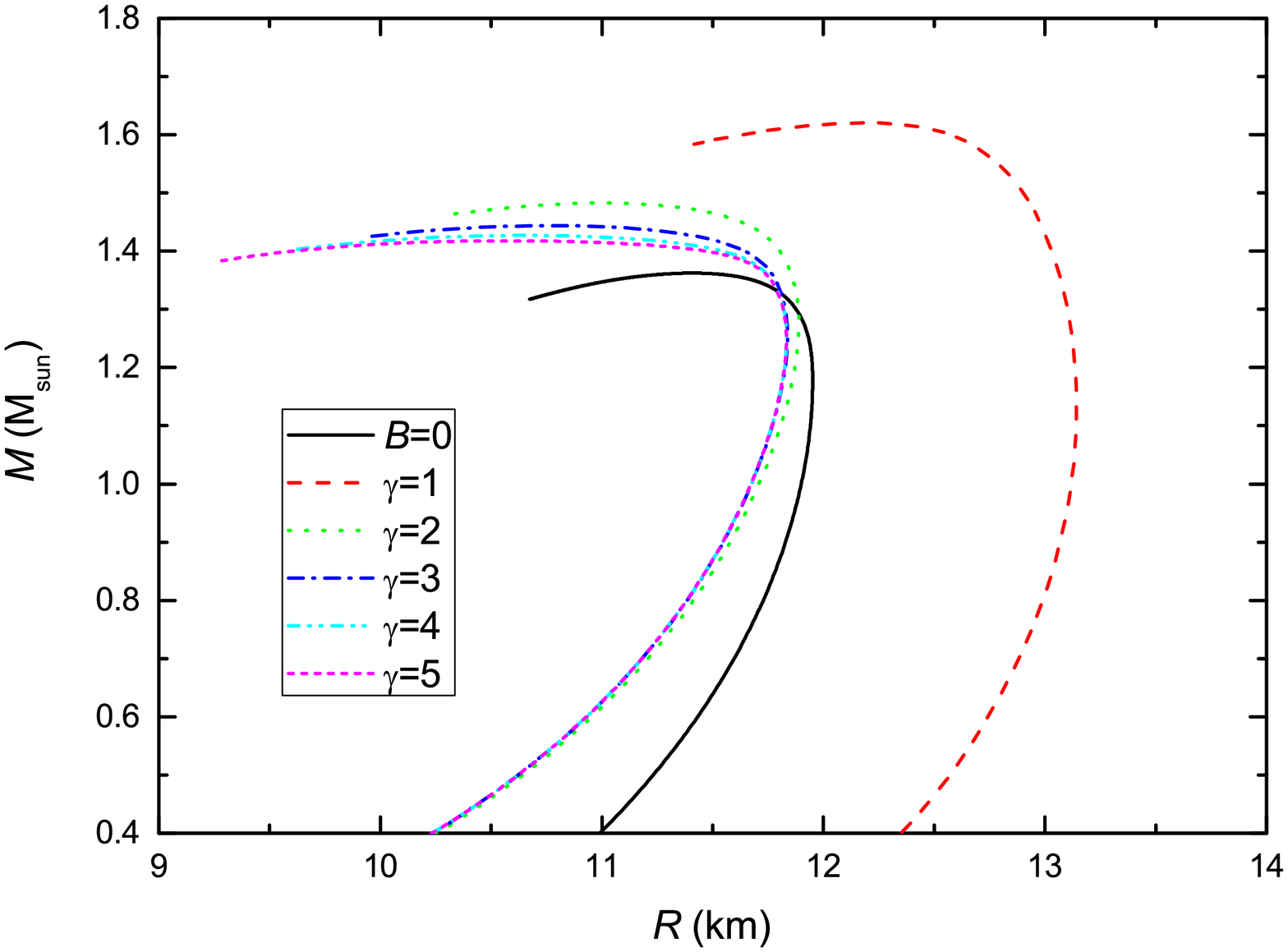}
\caption{(color online) Mass-radius relation for neutron stars without the magnetic field and with the magnetic field for different $\gamma$.}
\label{fig:m}
\end{figure}


\begin{table}[!h]
\centering
\caption{The maximum mass, corresponding radius and central energy density of neutron stars
for different magnetic fields with and without the AMMs. }
\begin{tabular}{c|ccc|ccc}
\hline
 &  \multicolumn{3}{c}{Not including the AMMs}   \vline & \multicolumn{3}{c}{Including the AMMs}   \\
\cline{2-7}  & $M$ (M$_{sun}$) & $R$ (km) & $\epsilon_c$ (fm$^{-4}$) & $M$ (M$_{sun}$) & $R$ (km) & $\epsilon_c$ (fm$^{-4}$) \\ \cline{2-7}
$B=0$ & 1.36 & 11.4 & 4.86  & 1.36 & 11.4 & 4.86 \\
$\gamma$=1 & 1.62 & 12.2 & 4.76  & 1.64 & 12.8 & 4.69 \\
$\gamma$=2 & 1.48 & 11.0 & 5.72  & 1.50 & 11.7 & 5.41 \\
$\gamma$=3 & 1.44 & 10.8 & 6.10  & 1.46 & 11.6 & 5.61 \\
$\gamma$=4 & 1.43 & 10.7 & 6.33  & 1.45 & 11.5 & 5.62 \\
$\gamma$=5 & 1.42 & 10.6 & 6.52  & 1.44 & 11.6 & 5.52 \\ \hline
\end{tabular}
\label{tab:max_mass}
\end{table}


The magnetic field  also alters the relative populations of particles  due to Landau quantization. In Fig. 3, we plot the relative populations as a function of baryon density, where $\rho_0$ is the nuclear saturation density. It can be seen that for $\rho < 2.5 \rho_0 $, the relative populations of proton, electron and muon are altered significantly by the magnetic field, while for larger density $\rho $, the change caused by the magnetic field is not as significant as in the low baryon density region.
With the increase of baryon density, the Fermi energy of particles becomes larger, and the ratio of the energy gap between adjacent energy levels of charged particles to their Fermi energy becomes smaller, which leads to a weaker influence of Landau quantization at high density.

\begin{figure}[!h]
\centering
\includegraphics[scale=0.4]{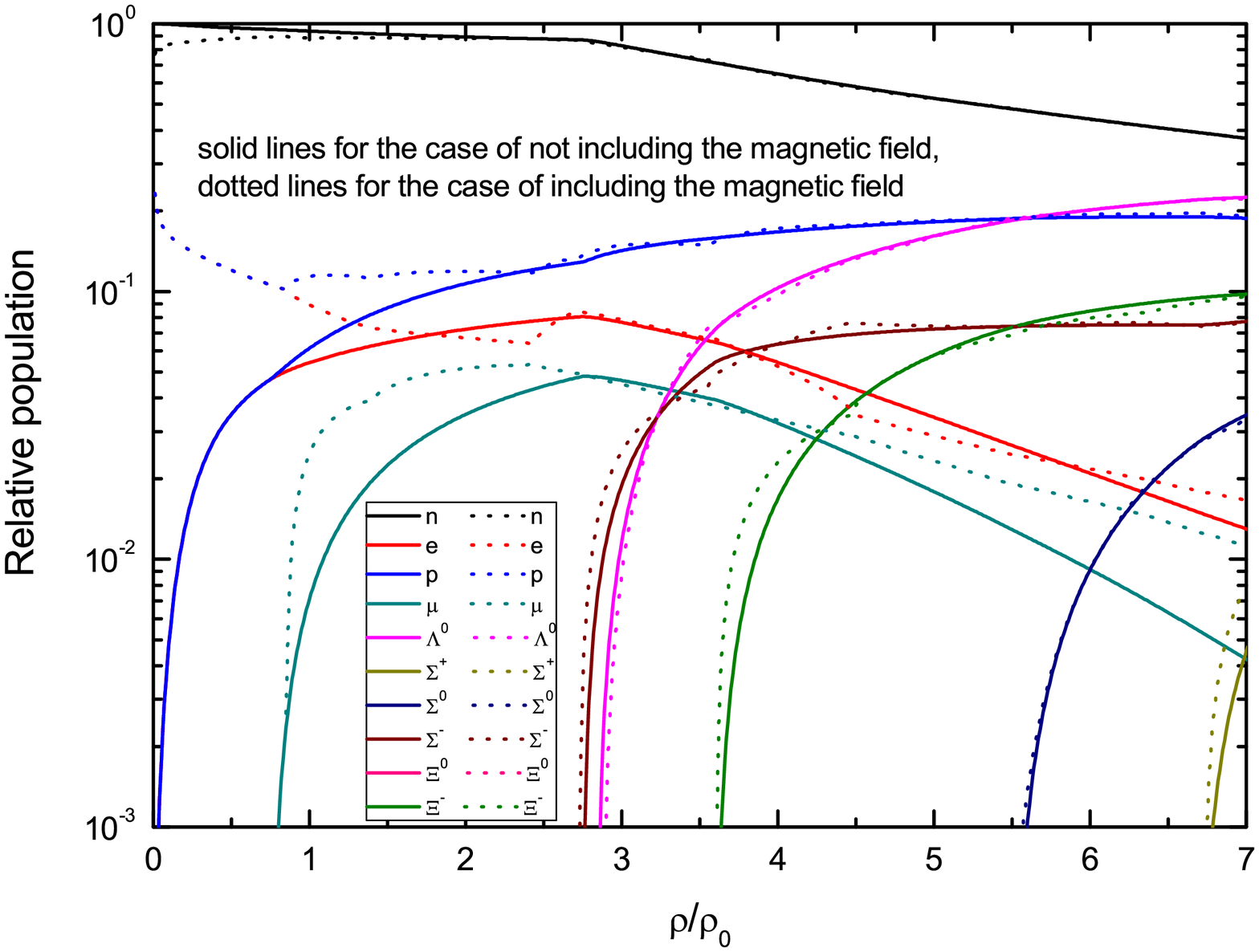}
\caption{(color online) Relative population of particles versus  baryon density. Solid lines for the case of not including the magnetic field,
dotted lines for the case of including the magnetic field. }
\label{fig:p}
\end{figure}

The magnetic field not only alters the relative populations of particles, but also alters the fractions of charged particles of the same kind  with different spin. In other words, the magnetic field causes charged particles to be spin-polarized. Obviously, in the absence of magnetic field, there will be no spin polarization. In Fig. 4, we plot the spin polarization as a function of baryon density in the presence of magnetic field. Two species of particles are selected to demonstrate the effect of the magnetic field, namely proton and neutron,
one electric charged, the other charge neutral. Neutron is not spin polarized. The magnetic moment decreases the energy of spin-up proton, but increases that of spin-down proton. So the spin polarization of proton is positive. For $\rho < 1.4 \rho_0$, the Fermi energy of proton is so low that all energy levels are occupied by spin-up protons. At higher baryon density, proton becomes less and less spin-polarized.

\begin{figure}[!h]
\centering
\includegraphics[scale=0.4]{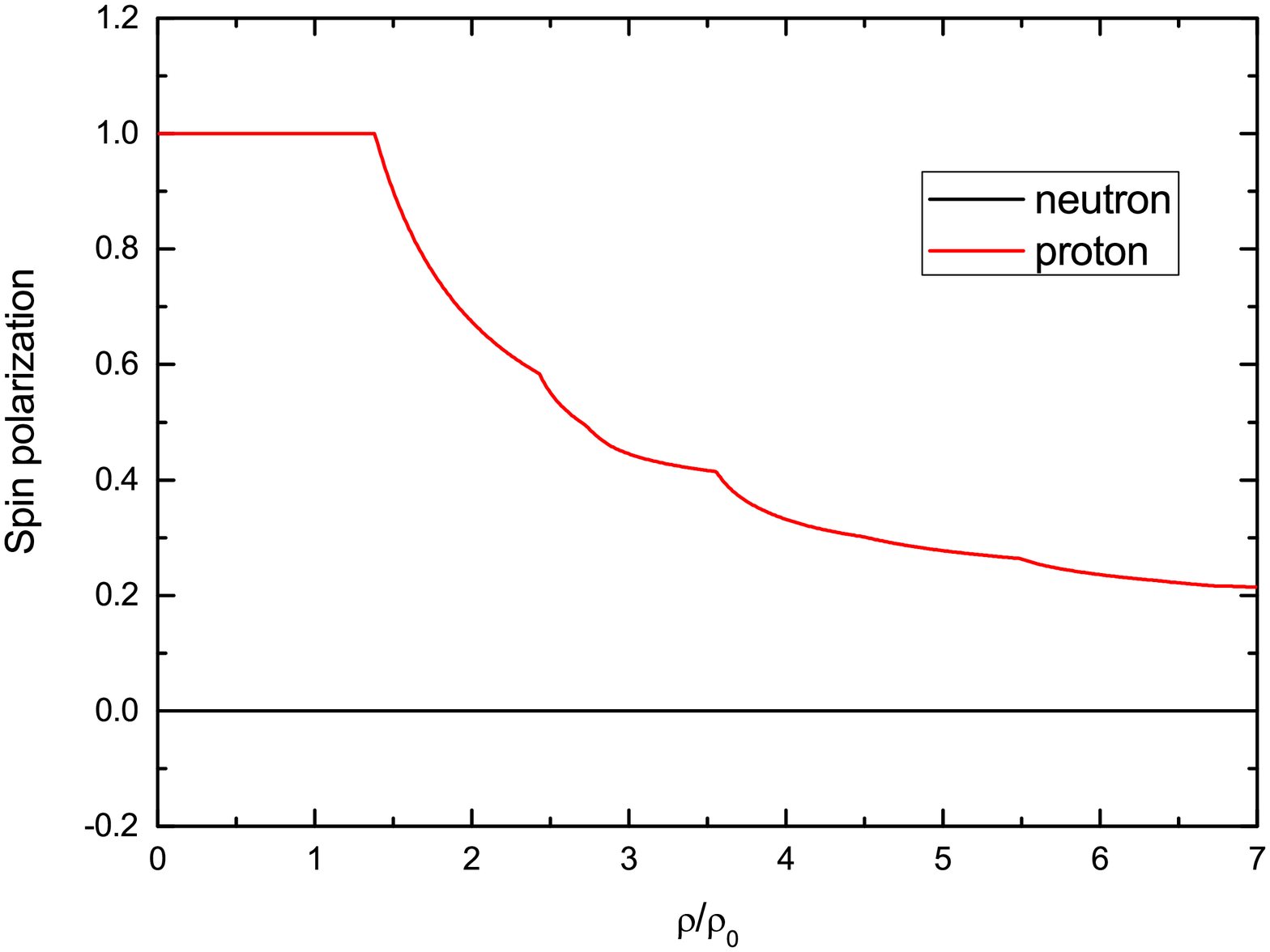}
\caption{(color online) Spin polarization of particles versus  baryon density in the presence of the magnetic field.}
\label{fig:sp}
\end{figure}

\subsection{The effect of the AMMs}

Now we are in a position to study the effect of the AMMs.
The AMMs alter the energy spectra of particles. Particles of the same kind  with different spin have different energy. So the AMMs influence
spin polarization of particles. In Fig. 5, spin polarization as a function of baryon density is displayed with and without the inclusion of the AMMs.
The most obvious difference between the two cases is that neutron is spin-polarized in the presence of magnetic field, due to the coupling of the AMM and the magnetic field. Because the  AMM of neutron is negative, its spin polarization is negative. The spin polarization of proton is increased by its AMM, since its AMM is positive. At lower baryon density, both proton and neutron are more spin-polarized because of their lower Fermi energy.

\begin{figure}[!h]
\centering
\includegraphics[scale=0.4]{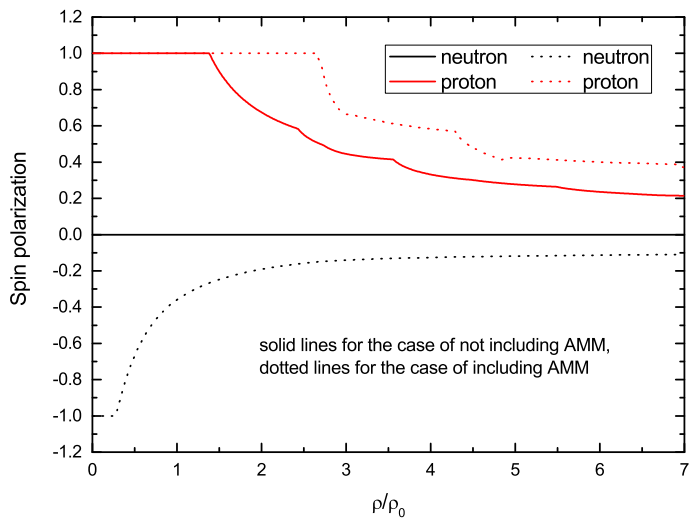}
\caption{(color online) Spin polarization of particles versus  baryon density with and without the AMMs. }
\label{fig:sp+a}
\end{figure}

The AMMs of particles also have an influence on the EOS of neutron star matter. In Fig. 6 we plot the EOS with and without the AMMs for $\gamma=3$. It can be seen
that the presence of the AMMs produces a small but not negligible  change to the EOS, and the AMMs do not always stiffen or soften the EOS.

\begin{figure}[!h]
\centering
\includegraphics[scale=0.4]{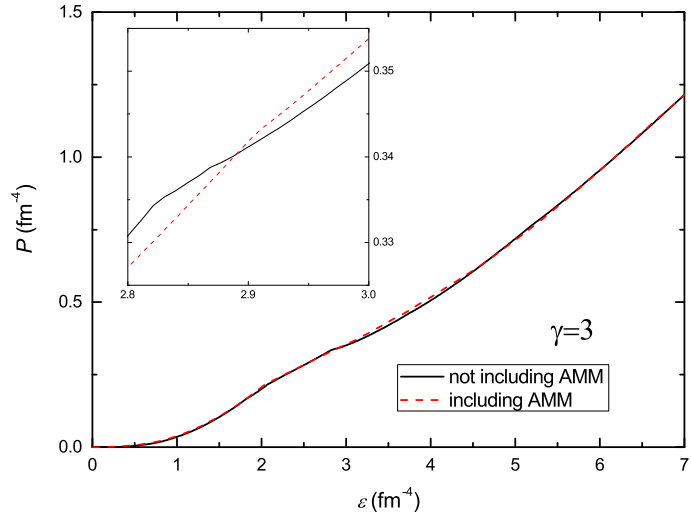}
\caption{ (color online) The EOS of neutron star matter with and without the AMMs. $\gamma =3$. }
\label{fig:s+a}
\end{figure}

The small change of EOS caused by the AMMs alters the mass-radius relation of neutron star. In Fig. 7 the mass-radius relation is plotted with and without the AMMs for $\gamma=1, 2, 3, 4, 5$. It can be seen the curves are shifted right by the AMMs. In the presence of the AMMs, neutron stars having the same mass will have a larger radius. The macroscopic properties of maximum mass neutron stars with and without
the AMMs can be seen from Table 3. For $1\leq\gamma\leq5$, the maximum mass is increased by 0.02 M$_{sun}$, while the corresponding radius is increased by around 0.8 km by the presence of the AMMs. The AMMs also decrease the central density of maximum mass neutron star.

\begin{figure}[!h]
\centering
\includegraphics[scale=0.4]{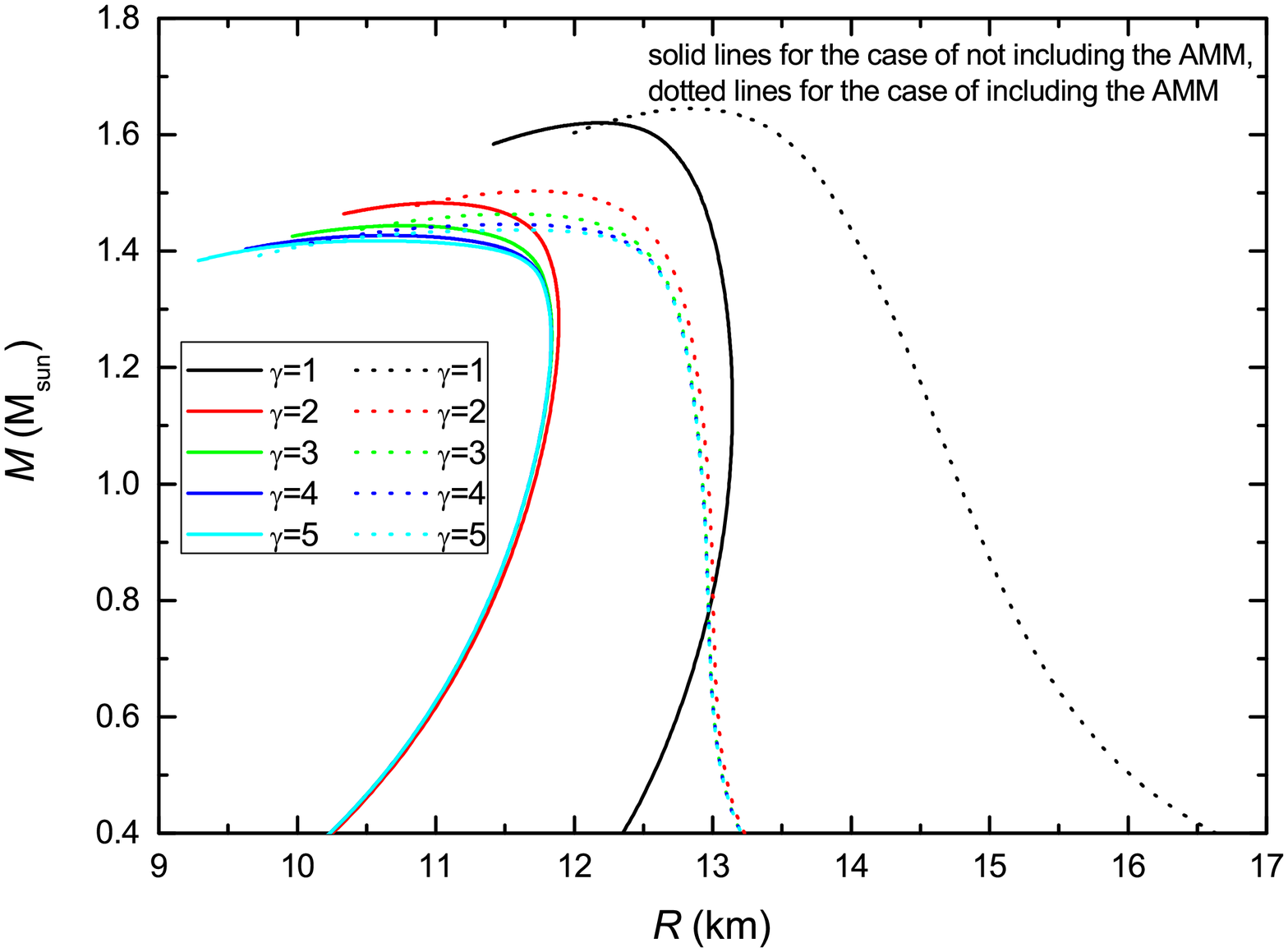}
\caption{(color online) Mass-radius relation for neutron stars with and without the AMMs.}
\label{fig:m+a}
\end{figure}

The AMMs also have an influence on the relative populations of particles. We plot the relative populations as a function of baryon density with and without the AMMs in Fig. 8. It can be seen that the AMMs alter the relative populations, but not significantly. The change does not follow a simple pattern.

\begin{figure}[!h]
\centering
\includegraphics[scale=0.4]{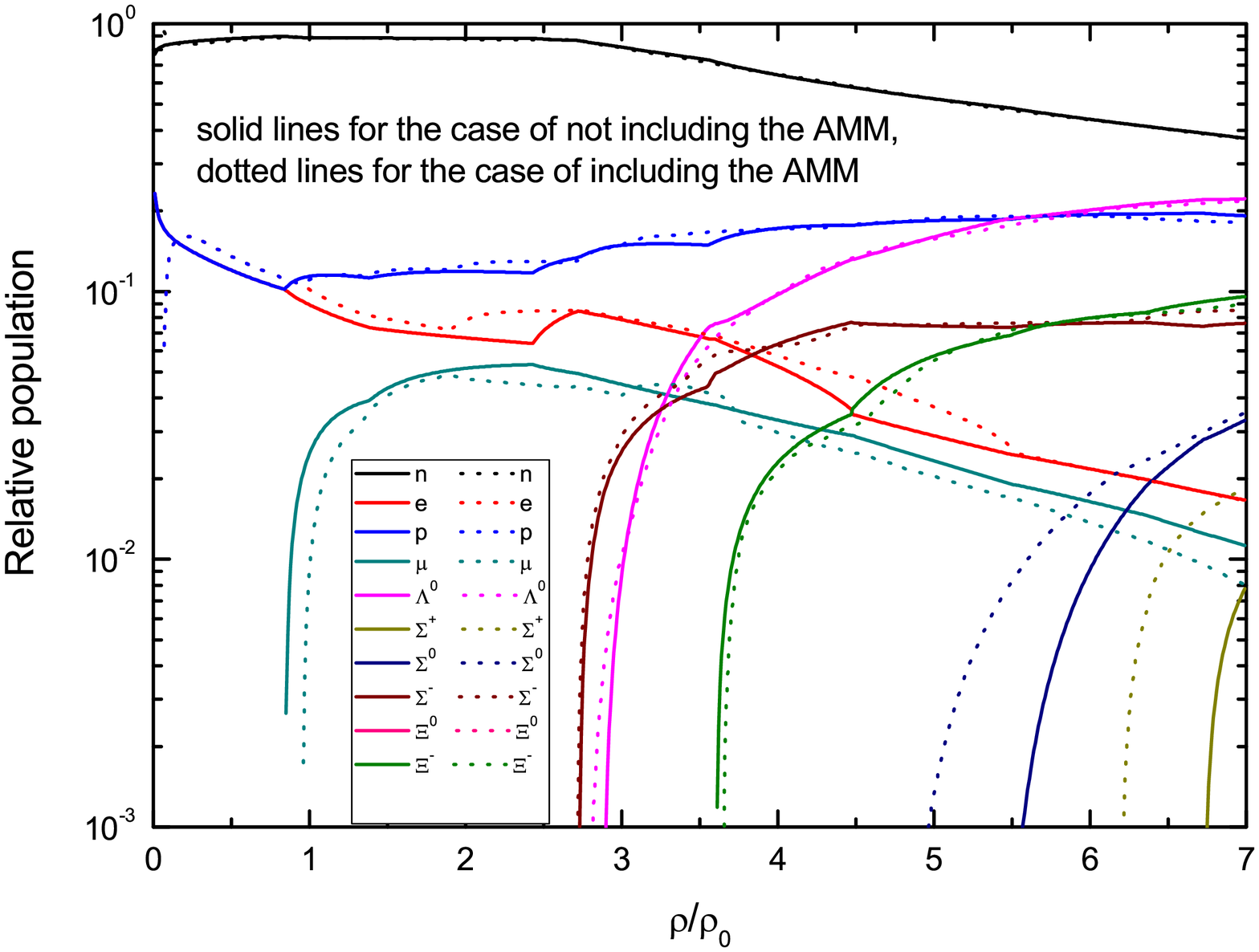}
\caption{(color online) Relative population of particles versus baryon density with and without the AMMs.}
\label{fig:p+a}
\end{figure}

\section{Summary and conclusion}
In this paper, we use the extended FSUGold model to study the properties of neutron stars with strong magnetic fields. The effect of the AMMs is also taken into account. Our theoretical framework has two major differences from most of previous works concerning the magnetic field.
One is that we use the chaotic field approximation, which is able to avoid the anisotropy problem. The other is that we use a energy density-dependent magnetic field model instead of a number density-dependent magnetic field model. We should point out that the maximum observed mass of neutron star is 2.01 M$_{sun}$, a fact that the extended FSUGold model is incapable of explaining. This situation could be remedied by modifying the high density behavior of the model, such as applying the recently proposed $\sigma$
-cut scheme \cite{Maslov:2015lma}.

It is found that the magnetic field stiffens the EOS. For $2\leq\gamma\leq5$, it increases the maximum mass of neutron star by about 6\%. For baryon density not too large, it alters the relative populations of particles considerably. Charged fermions are spin polarized by the presence of the magnetic field.

The AMMs only have a small influence on the EOS, and increases the maximum mass of neutron star by about 0.02 M$_{sun}$.
The corresponding radius is increased by about 0.8 km. In addition, neutral fermions are spin polarized by the presence of the AMMs.
\\

\section*{Acknowledgments}
The authors wish to thank Dr. Ren-li Xu  and Prof. Xu-guang Huang for useful correspondences.
This work was supported by the National Natural
Science Foundation of China (Grant Nos. 11535004,
11375086, 11120101005, 11175085, and 11235001), by the
973 National Major State Basic Research and Development
of China, Grant No. 2013CB834400 and by the
Science and Technology Development Fund of Macau under
Grant No. 068/2011/A.

\vspace{-1mm}
\centerline{\rule{80mm}{0.1pt}}
\vspace{2mm}



\begin{thebibliography}{90}

\vspace{3mm}


\bibitem{brito2014superradiant} R. Brito, V. Cardoso, and P. Pani, Phys. Rev. D, {\bf 89}: 104045 (2014)

\bibitem{konoplya2008quasinormal} R. A. Konoplya and R. D. B. Fontana, Phys. Lett. B, {\bf 659}: 375---379 (2008)

\bibitem{wu2015decay} C. Wu and R. L. Xu, Eur. Phys. J. C, {\bf 75}: 391 (2015)

\bibitem{duncan1992formation} R. C. Duncan and C. Thompson, Astrophys. J., {\bf 392}: L9 (1992)

\bibitem{thompson1995soft} C. Thompson and R. C. Duncan, Mon. Not. Roy. Astron. Soc., {\bf 275}: 255---300 (1995)

\bibitem{usov1992millisecond} V. V. Usov, Nature, {\bf 357}: 472---474 (1992)

\bibitem{1538-4357-486-2-L129} G. Vasisht and E. V. Gotthelf, Astrophys. J., {\bf 486}: L129 (1997)

\bibitem{1538-4357-519-2-L139} P. M. Woods, C. Kouveliotou, J. van Paradijs, et al., Astrophys. J., {\bf 519}: L139 (1999)

\bibitem{kouveliotou1998x} C. Kouveliotou, Nature, {\bf 393}: 235---237 (1998)

\bibitem{broderick2000equation} A. Broderick, M. Prakash, and J. M. Lattimer, Astrophys. J., {\bf 537}: 351 (2000)

\bibitem{oppenheimer1939massive} J. R. Oppenheimer and G. M. Volkoff, Phys. Rev., {\bf 55}: 374---381 (1939)

\bibitem{wu2011strange} C. Wu and Z. Z. Ren, Phys. Rev. C, {\bf 83}: 025805 (2011)

\bibitem{lopes2015magnetized} L. L. Lopes and D. Menezes, JCAP, {\bf 1508}: 002 (2015)

\bibitem{broderick2002effects} A. E. Broderick, M. Prakash, and J. M. Lattimer, Phys. Lett. B, {\bf 531}: 167---174 (2002)

\bibitem{bandyopadhyay1997quantizing} D. Bandyopadhyay, S. Chakrabarty, and S. Pal, Phys. Rev. Lett., {\bf 79}: 2176---2179 (1997)

\bibitem{rabhi2009quark} A. Rabhi, H. Pais, P. K. Panda, et al., J. Phys. G, {\bf 36}: 115204 (2009)

\bibitem{menezes2009quark} D. P. Menezes, M. Benghi Pinto, S. S. Avancini, et al., Phys. Rev. C, {\bf 80}: 065805 (2009)

\bibitem{ryu2010medium} C. Y. Ryu, K. S. Kim, and M.-K. Cheoun, Phys. Rev. C, {\bf 82}:025804 (2010)

\bibitem{rabhi2011warm} A. Rabhi, P. K. Panda, and C. Providencia, Phys. Rev. C, {\bf 84}: 035803 (2011)

\bibitem{mallick2011possible} R. Mallick and M. Sinha, Mon. Not. Roy. Astron. Soc., {\bf 414}: 2702---2708 (2011)

\bibitem{lopes2012influence} L. L. Lopes and D. P. Menezes, Braz. J. Phys., {\bf 42}: 428---436 (2012)

\bibitem{casali2014hadronic} R. H. Casali, L. B. Castro, and D. P. Menezes, Phys. Rev. C, {\bf 89}: 015805 (2014)

\bibitem{menezes2014repulsive} D. P. Menezes, M. B. Pinto, L. B. Castro, et al., Phys. Rev. C, {\bf 89}: 055207 (2014)

\bibitem{mallick2014deformation} R. M. S. Schramm and S. Schramm, Phys. Rev. C, {\bf 89}: 045805 (2014)

\bibitem{Yuan}
Y. F. Yuan and J. L. Zhang, Astrophys. J., {\bf 525}: 950 (1996)

\bibitem{todd2005neutron} B. G. Todd-Rutel and J. Piekarewicz, Phys. Rev. Lett., {\bf 95}: 122501 (2005)

\bibitem{piekarewicz2007validating} J. Piekarewicz, Phys. Rev. C, {\bf 76}: 064310 (2007)

\bibitem{fattoyev2010sensitivity} F. J. Fattoyev and J. Piekarewicz, Phys. Rev. C, {\bf 82}: 025810 (2010)

\bibitem{glendenning1982hyperon} N. K. Glendenning, Phys. Lett. B, {\bf 114}: 392---396 (1982)

\bibitem{mao2003study} G. J. Mao, A. Iwamoto, and Z. X. Li, Chin. J. Astron. Astrophys., {\bf 3}: 359---374 (2003)

\bibitem{wu2013neutron} C. Wu, W. L. Qian, Y. G. Ma, et al., Int. J. Mod. Phys. E, {\bf 22}: 1350026 (2013)

\bibitem{xu2014superfluidity} R. L. Xu, C. Wu, and Z. Z. Ren, Int. J. Mod. Phys. E, {\bf 23}: 1450078 (2014)

\bibitem{Chakrabarty:1996te}
  S. Chakrabarty, Phys. Rev. D {\bf 54}: 1306 (1996)

\bibitem{strickland2012bulk} M. Strickland, V. Dexheimer, and D. P. Menezes, Phys. Rev. D, {\bf 86}: 125032 (2012)

\bibitem{baldo1997microscopic} M. Baldo, I. Bombaci, and G. F. Burgio, Astron. Astrophys., {\bf 328}: 274---282 (1997)

\bibitem{kalogera1996maximum} V. Kalogera and G. Baym, Astrophys. J., {\bf 470}: L61---L64 (1996)

\bibitem{dexheimer2012hybrid} V. Dexheimer, R. Negreiros, and S. Schramm, Eur. Phys. J. A, {\bf 48}: 189 (2012)

\bibitem{dong2013dense} J. M. Dong, U. Lombardo, W. Zuo et al, Nucl. Phys. A, {\bf 898}: 32---42 (2013)

\bibitem{gomes2014effects} R. O. Gomes, V. Dexheimer, and C. A. Z. Vasconcellos, Astron. Nachr., {\bf 335}: 666 (2014)

\bibitem{sinha2013hypernuclear} M. Sinha, B. Mukhopadhyay, and A. Sedrakian, Nucl. Phys. A, {\bf 898}: 43---58 (2013)

\bibitem{Maslov:2015lma}
  K. A. Maslov, E. E. Kolomeitsev, and D.N.Voskresensky, Phys. Rev. C, {\bf 92}: 052801 (2015)





\end{thebibliography}


\end{CJK*}
\end{document}